\documentclass[12pt,english]{article}
\pdfoutput=1
\usepackage[T1]{fontenc}
\usepackage[latin1]{inputenc}
\usepackage{geometry}
\usepackage{graphicx}
\usepackage{setspace}
\makeatletter

\usepackage{graphicx}

\usepackage[round, comma, sort&compress, authoryear]{natbib}

\usepackage{xspace}

\newcommand{\pdb}{\textsc{pdb}\xspace}
\newcommand{\rna}{\textsc{rna}\xspace}
\newcommand{\dna}{\textsc{dna}\xspace}
\newcommand{\gabb}{\textsc{gabb}\xspace}
\newcommand{\rmsd}{\textsc{rmsd}\xspace}

\newcommand{\rapper}{\textsc{rapper}\xspace}
\newcommand{\rappertk}{\textit{Rapper}tk\xspace}
\newcommand{\mdsa}{\textsc{mdsa}\xspace}
\newcommand{\cns}{\textsc{cns}\xspace}

\newcommand{\Ang}[1]{${#1}$\AA\xspace}

\date{}

\usepackage{babel}
\makeatother
\begin{document}

\title{\rna sampling and crystallographic refinement using \rappertk%
\thanks{This document is very similar to a chapter in SG's thesis submitted
in Sept.2007 to the University of Cambridge, England.%
}}

\author{Swanand Gore and Tom Blundell\\
\{swanand,tom\}@cryst.bioc.cam.ac.uk\\
Department of Biochemistry, University of Cambridge\\
Cambridge CB2 1GA England}

\maketitle
\begin{abstract}
\textbf{Background} Dramatic increases in \rna~ structural data
have made it possible to recognize its conformational preferences
much better than a decade ago. This has created an opportunity to
use discrete restraint-based conformational sampling for modelling
\rna~ and automating its crystallographic refinement.

\textbf{Results} All-atom sampling of entire \rna~ chains, termini
and loops is achieved using the Richardson \rna~ backbone rotamer
library and an unbiased distribution for glycosidic dihedral angle.
Sampling behaviour of \rappertk~ on a diverse dataset of \rna~
chains under varying spatial restraints is benchmarked. The iterative
composite crystallographic refinement protocol developed here is demonstrated
to outperform \cns-only refinement on parts of t\textsc{rna}$^{Asp}$
structure.

\textbf{Conclusion} This work opens exciting possibilities for further
work in \rna~ modelling and crystallography.
\end{abstract}

\newpage
\section{Introduction}

\subsection{Role of \rna}

\rna is involved in many important biochemical functions involving
genetic information, such as its storage (viral \rna), communication
(m\rna) and modulation (sno\rna, micro\rna). \rna also performs
protein-like functions like enzymatic catalysis (ribosomal peptide
bond formation - r\rna) and specific binding (amino-acid-specific
t\rna) etc. It is believed to have played a major role in the early
evolution of cellular life because it is functionally intermediate
to proteins and \dna, exhibiting enzymatic activity as well as information
storage and transfer (\cite{VoetVoet}). There is an increasing recognition
of \rna's importance in cellular life (\cite{SchlickRNAchapter})
and attempts to organize available experimental information as \rna
ontology (\cite{RNAontoCons}).

\subsection{\rna structure}

\rna is simpler than proteins in the sequence space due to a much
smaller alphabet, but structurally it is more complicated. A typical
nucleotide contains at least thrice as many non-hydrogen atoms as
an amino acid residue. The most prominent parts of polynucleotide
structures are nucleotide bases which are purines or pyrimidines.
Purines adenosine and guanine are 5,6 aromatic rings and resemble
tryptophan's sidechain. Pyrimidines uracyl and cytosine are aromatic
6-rings which resemble phenylalanine and tyrosine sidechains. The
bases can undergo a variety of post-transcriptional modifications,
increasing the effective number of base types (\cite{RNAmodomics}).
A striking feature of \dna and (most) \rna structures is the common
Watson-Crick pairing of purines with pyrimidines, and the associated
base stacking. But in \rna structures, there are other non-canonical
base interactions which contribute to stabilization of various \rna
motifs (\cite{RNAmotifsReview}). Bases are linked to 5-membered ribose
sugar rings through glycosidic linkages. The $\chi$ torsion angle,
which describes base rotation with respect to sugar, is distributed
around $-120^{o}$ or diametrically opposite to it, around $-60^{o}$
(\cite{RNAconfClasses}). Sugar ring connects bases to the backbone,
and occurs only in two conformations, $C3'$-endo or $C2'$-endo.
The phosphate-sugar backbone has six torsion angles ($\alpha,\beta,\gamma,\delta,\epsilon,\zeta$)
and much greater freedom than the protein mainchain. But conformational
correlations in that space have been recognized recently (\cite{RNA2virtualdihed},
\cite{RNAlib}, \cite{RNAconfClasses}).

Despite chemical differences, protein and \rna chains are logically
similar. \rna backbone and protein mainchain are the unbranched chains
in both polymers and show clear preferences for parts of their dihedral
spaces. In proteins, mainchain completely determines $C_{\beta}$
coordinate and similarly, \rna backbone almost completely determines
the sugar coordinates. Bases are similar to sidechains, because both
are rotameric and confer chemical characteristics to respective polymers%
\footnote{But base rotamericity is weaker and not used in this work.%
}. Thus, \rna backbone, sugar and bases are analogous respectively
to protein mainchain, $C_{\beta}$ atom and sidechains.

\subsection{\rna structure prediction}

Like other biopolymers, sequence data for \rna is far greater than
3D structural data. \rna crystals generally do not diffract as well
as proteins because \rna is harder to purify and crystallize, possibly
due to size and flexibility. Hence structure prediction methods are
important to bridge the sequence-structure gap. \rna structure prediction
is done at two levels - secondary and 3D. Secondary structure prediction
is important because it can help identify a variety of motifs like
stem, hairpin loop, internal loop, junction loop, bulges and pseudoknots.
These predictions can prove to be important restraints to guide further
3D structure prediction. 3D structure prediction is important to locate
interesting sites and tertiary interactions, but it has so far been
dependent on secondary structure prediction (\cite{RNAstrpredReview}).

Secondary structure prediction estimates the base pairings given a
sequence. Due to standard Watson-Crick base-pairing, \rna commonly
exhibits helical stem regions. The sequence that connects the two
strands in a stem is called a loop. Stem and loop arrangement can
develop in a hierarchical fashion, giving rise to a structure that
can be represented like a tree. Dynamic programming based algorithms
like Mfold (\cite{mfold}), Sfold (\cite{RNAfold}), RNAstructure
(\cite{RNAstructure}) assign secondary structure in such a way as
to minimize the free energy for the sequence%
\footnote{Free energies used here are experimentally determined as a function
of host secondary structure type and base-pairing.%
}. Optimal and highly-ranked suboptimal solutions are very likely to
contain the correct secondary structure. Suboptimal solutions can
be filtered using Boltzmann sampling (\cite{sfold}) or abstract shape
analysis (\cite{RNAshapes}) to enrich the solutions of dynamic programming
algorithms. In addition to dynamic programming, various other approaches
have also been utilized such as genetic algorithms (\cite{MPGAfold})
and Monte-Carlo sampling (\cite{kinefold}). All approaches can be
further enhanced by using multiple sequence alignments, based on the
information-theoretic principle that MSAs improve the signal to noise
ratio.

Tree-like simplicity of \rna secondary structure is lost when pseudoloops
are formed by base-pairing of a stretch in loop with another strand.
Pseudoloops are known to occur in many more complicated ways than
the simplest H-type. They reduce flexibility of the structure because
often the stems involved in a pseudoloop are coaxially stacked. Dynamic
programming algorithms which include general psudoknots scale poorly
but simple H-type pseudoknots can be incorporated without loss of
efficiency (\cite{RNAstrpredReview}).

Fully-automated 3D structure prediction procedures are yet to be devised
for \rna. This is perhaps due to the complexity of \rna structure
and relatively less structural information as \rna is studied more
often from a non-structural perspective. Present approaches encoded
in programs like E\rna-3D (\cite{ERNA3d}), \textsc{rna}2D3D (\cite{RNA2d3d})
and S2S (\cite{s2s}) are focussed on assisting the 3D model building
exercise interactively. The inputs are a combination of known/predicted
secondary structure, features derived from 3D structural data and
available experimental restraints. The interactively assembled model
is generally subjected to molecular dynamics refinement and minimization
(\cite{RNAstrpredReview}).

Recurrent 3D motifs in \rna structure are short sequence-dependent
combinations of backbone conformations and base interactions. A complex
set of noncovalent interactions stabilize them. Motif identification
has not matured enough to be usable in 3D structure prediction (\cite{RNAmotifsReview}).

\subsection{\rna crystallographic refinement}

\rna crystallography is harder than protein crystallography because
nucleotides are bigger and more flexible than amino acid residues.
\rna crystals rarely diffract better than \Ang{2}. Due to many high-quality
protein structures, their statistical preferences can be used effectively
to solve more protein structures. This critical mass effect is yet
to be achieved for \rna as there are not enough structures for confident
identification of backbone preferences and 3D motifs. Apart from their
stand-alone utility, high-quality single-chain \rna structures are
also essential for docking into low-resolution EM data of large complexes
containing \rna chains.

Temperature factors suggest that \rna flexibility is the least for
paired bases and the highest for phosphates. Yet phosphates are also
easy to detect due to greater electron density. Hence \rna crystallographer
identifies bases and phosphates of \rna chain in the initial map
and then iteratively completes and refines the structure. Due to lack
of structural preferences, this process is manual, tedious and laborious.
Methods and progress in \rna crystallography have been reviewed by
\cite{RNAcrystReview} and \cite{RNAlongshort}.

\subsection{This work}

This work is inspired by the success of \rapper's protein sampling
which proved effective in loop sampling, comparative modelling and
automation of crystallographic refinement (\cite{RAPPER_amber}, \cite{rapperKnowledgeXray},
\cite{rapperLowResolution}). It is the last task that would be very
useful to the crystallographer if replicated for \rna. In protein
crystallography, approximate locations of $C_{\alpha}$ atoms and
sidechains identified by the crystallographer are sufficient for \rapper
to reach an almost refined structure. It is expected that a similar
approach would work for \rna chains too, given the approximate locations
of phosphates and bases visually identifiable in the electron density.
Apart from crystallographic use, a generalized restraint-based all-atom
sampler of \rna would be useful for generating decoy structures useful
for benchmarking of energy functions. It would also allow generation
of models with a prescribed sequence and secondary structure, and
serve as a tool for generating 3D models of \rna motifs.

In this work, we show that \rapper's \gabb (genetic algorithm using
branch-and-bound technique) algorithm can be extended to \rna structures
to sample it accurately and efficiently under a variety of positional
restraints on backbone and bases. We also demonstrate the all-atom
iterative crystallographic refinement of parts of a t\textsc{rna}$^{Asp}$
structure.

\section{\rna tracing}

These benchmarks assess the utility of \rna sampling for the intended
application of crystallographic refinement, hence the restraints chosen
here reflect the kind of information a crystallographer can provide.
Spherical positional restraints are used for phosphates ($P$ atoms)
and $C4'$ atoms. Base planes are restrainted using a union-of-spheres
restraint (baseplane restraint). This restraint is satisfied when
the sampled set of coordinates lie within the union of given spheres.

As described in \cite{rappertk}, \rappertk uses the Richardson rotamer
library (\cite{RNAlib}) for \rna backbone sampling. Sugar-phosphate
backbone consists of six dihedral angles : $\alpha$ ($O3'_{i-1}-P_{i}-O5'_{i}-C5'_{i}$),
$\beta$ ($P_{i}-O5'_{i}-C5'_{i}-C4'_{i}$), $\gamma$ ($O5'_{i}-C5'_{i}-C4'_{i}-C3'_{i}$),
$\delta$ ($C5'_{i}-C4'_{i}-C3'_{i}-O3'_{i}$), $\epsilon$ ($C4'_{i}-C3'_{i}-O3'_{i}-P_{i+1}$)
and $\zeta$ ($C3'_{i}-O3'_{i}-P_{i+1}-O5'_{i+1}$). \cite{RNAlib}
define the \rna backbone suite as a set of seven dihedral angles
$\{\delta^{i},\epsilon^{i},\zeta^{i},\alpha^{i+1},\beta^{i+1},\gamma^{i+1},\delta^{i+1}\}$
and identify $42$ distinct rotamers. In a recent effort, this library
has been extended to $46$ rotamers, with standard deviations specified
for each cluster (J. M. Richardson, personal communication). Glycosidic
dihedral $\chi$ is defined over $O4'_{i}-C1'_{i}-N1'_{i}-C2'_{i}$
for pyrimidines $O4'_{i}-C1'_{i}-N9'_{i}-C4'_{i}$ for purines. $\chi$
preferences have not been rigorously analyzed, hence in this work
it is randomly sampled between $-180^{o}$ and $+180^{o}$ in steps
of $10^{o}$. 

The basic operation in \rna chain extension is building next or previous
backbone suite by sampling a backbone suite rotamer and building sugar/base
of present nucleotide using a random sample for glycosidic linkage.
Various styles of sampling use this building block in different ways.

\subsection{Sampling styles}

\begin{figure}

\caption{Reverse RNA builder}

\begin{center}\includegraphics[%
  width=100mm]{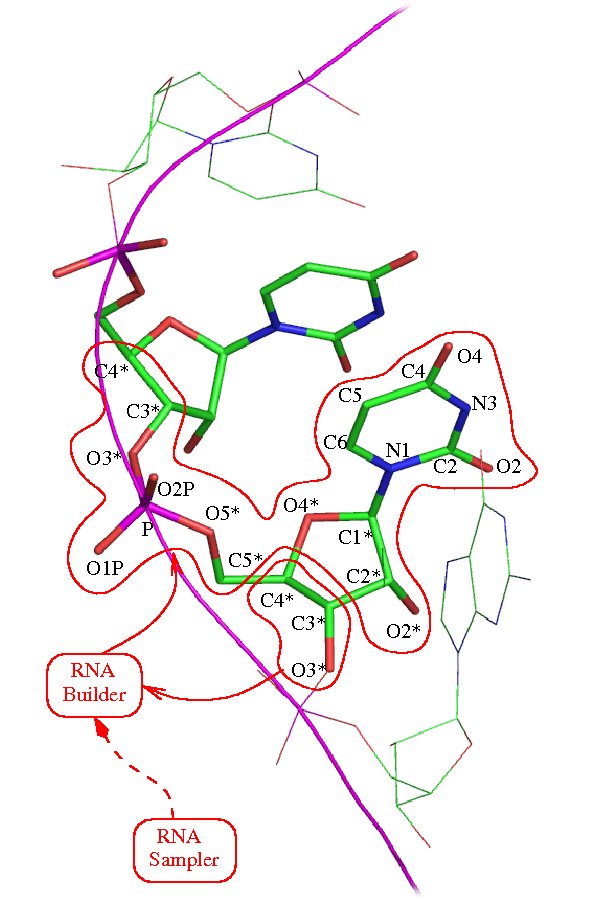}\end{center}

\label{revRNAbuilder}
\end{figure}

For iterative crystallographic refinement, basic operations over the
\rna chain are rebuilding the whole chain, or its terminal ($5'$
or $3'$) or an intermediate fragment (loop) :

\begin{itemize}
\item Forward sampling ($5'\to3'$) is performed using the default \rna
builder as described in \cite{rappertk}. This builder depends on
atoms $(C5'_{i},C4'_{i},C3'_{i})$ and yields $(O3'_{i},P_{i+1},$
$O5'_{i+1},C5'_{i+1},$ $C4'_{i+1},C3'_{i+1})$ atoms (see \cite{rappertk} for
figure). It also builds the sugar and base of $i^{th}$ nucleotide.
\item Bootstrapping required for sampling the whole chain is explained in
\cite{rappertk}. It involves approximate positioning of $(P,O1P,O2P,O5',C5',C4',C3')$
atoms of the first nucleotide.
\item Backward sampling ($3'\to5'$) is performed by slightly changing the
forward builder. The same backbone rotamers are sampled, but the builder
depends on atoms $(O3'_{i},C3'_{i},C4'_{i})$ to calculate coordinates
for atoms $(C5'_{i},O5'_{i},P_{i-1},O3'_{i-1},C3'_{i-1},C4'_{i-1})$
(see Fig.\ref{revRNAbuilder}). Sugar and base for $i^{th}$ nucleotide
are also built.
\item Loop sampling uses forward sampling. Nucleotides between and including
$start$ and $end$ indices are rebuilt. Base of $(start-1)^{th}$
nucleotide is resampled within \Ang{2} positional restraints. Approximate
loop closure is achieved by partial sampling of $(end+1)^{th}$ nucleotide's
$(P,C5',C4',C3')$ atoms under similar restraints. Loop closure restraint
is back-propagated by enforcing a spherical positional restraint centered
at $P^{end+1}$ atom with radius $7*(end+1-i)$ \Ang{} on $P^{i}$
atom and also forcing it to remain \Ang{5} away from the $P^{end+1}$
atom.
\end{itemize}

\begin{table}

\caption{Dataset of RNA chains used for the tracing exercise.}

\begin{center}\begin{tabular}{|c|c|c||c|c|c|}  

\hline

 PDB id & Size$^a$ & Filtered Size$^b$&

PDB id & Size$^a$ & Filtered Size$^b$\\\hline\hline

 1i6u&  37&   11& 1jj2& 121&   18\\\hline  1kh6&  27&   10& 1l9a& 125&   15\\\hline  1l2x&  27&   10& 1n78&  75&   24\\\hline  2fmt&  77&   17& 361d&  19&    1\\\hline  1mzp&  55&   18& 1k8w&  21&   10\\\hline  1duh&  44&   19& 1kq2&   6&    0\\\hline  1cx0&  71&   18& 1f7y&  56&    9\\\hline  1ec6&  19&    6& 1c0a&  77&   37\\\hline  1m5k&  91&   20& 1kxk&  69&   11\\\hline  1b7f&  12&    3& 1hq1&  48&   35\\\hline  1f1t&  37&    4& 1ivs&  75&    6\\\hline  1cvj&   8&    1& 1gid& 158&   29\\\hline  1b23&  73&   13& 1qtq&  73&   28\\\hline  1ddy&  35&    4& 1lng&  96&    8\\\hline  1e7x&  16&   16& 1f27&  18&   11\\\hline  1et4&  35&    9& 1jbs&  28&   11\\\hline  1g1x&  39&   14& 1ehz&  76&   26\\\hline  1f7u&  75&   23& 1hr2& 156&   19\\\hline  1hmh&  34&   11& 1mji&  33&    7\\\hline  1jbt&  28&    7& 1ffy&  74&   11\\\hline  1qf6&  76&   25& 1e7k&   9&    4\\\hline  1m8x&   8&    3& 1ntb&  21&    4\\\hline  1ddl&   7&    0& 1ser&  64&    9\\\hline  1h4s&  67&   12& 429d&  12&    6\\\hline  \end{tabular}  \end{center}

$^{a}$Size is the number of nucleotides in the chain as in the deposited
PDB structure.

$^{b}$Filtered size is the size of the largest contiguous segment
of rotameric backbone suites in the given chain. The Richardson RNA
backbone rotamer set consists of $46$ 7-dihedral tuples, along with
standard deviations for all dihedrals in each tuple. A backbone suite
is rotameric if the largest single-dihedral difference between the
suite and the closest Richardson backbone rotamer is $<30^{o}$ or
$<3\sigma$ of that dihedral angle.

\label{rnaTraceDataset}
\end{table}

\subsection{Initial observations}

For benchmarking of \rna tracing capabilities, we have used a set
of diverse \rna chains compiled by \cite{RNA2virtualdihed} for their
virtual dihedral analysis (summarized in Table \ref{rnaTraceDataset}).
In the first exercise, we restrained $P,C4'$ atoms to \Ang{2} positional
restraints and sampled only the backbones of the chains. But a model
could be generated for only $12$ of the $48$ chains. This suggested
that the Richardson rotamer set consisting of only $46$ states was
too coarse-grained for the task being attempted. Indeed, $46$ is
a small number for capturing preferences of a flexible backbone consisting
of $6$ dihedral angles. Hence it was decided to supplement sampling
by perturbation - after a rotamer is sampled, a random noise within
$1\sigma$ of the respective dihedrals is added to them. Standard
deviations were kindly provided by J. M. Richardson (personal communication).
When the latest exercise was repeated with perturbed sampling, at
least one model could be generated for $47$ of $48$ examples. When
perturbed backbone-only sampling was performed within tighter \Ang{1}
restraints on $P,C4'$ atoms, this dropped to $36$ of $48$ chains.
These failures could be traced to the non-rotameric backbone suites
present in the chains. Then the longest stretch of good backbone suites
was identified within every chain. A good suite was defined to be
the one for which the largest single angle difference from the closest
rotamer was within $30^{o}$ or $3\sigma$ for that angle. As seen
from Table \ref{rnaTraceDataset}, such good fragments are fairly
small as compared to whole chain. $45$ of $46$ such fragments could
be sampled successfully under the same restraints. On these fragments,
all-atom sampling was also possible within \Ang{1} restraints on
$P,C4'$ atoms and \Ang{5} baseplane restraints. By dropping the
$C4'$ from these restraints, an increase in sampling time was observed,
accompanied by a reduction in number of examples for which 10 models
could be built ($33$ of $45$). This is due to population dilution,
which in this case is the reduction in number of members which will
satisfy restraints for the base of next nucleotide. As expected, using
stricter base restraint of \Ang{3} made the matters worse due to
greater base restraint violations and no propagation of base restraints
onto the backbone. Base restraint used here is hard to satisfy closely
because a small error close to sugar amplifies towards the far end
of the planar base. This problem can be addressed if given base restraint
can be propagated onto $C4'$ atom, but it is unclear at present how
to achieve this.

\subsection{Sampling performance}

Two characteristics are desirable in a sampling process: (a) given
tight restraints, sampling should be efficient and (b) given loose
restraints, sampling should produce native-like conformations owing
to the knowledge of native structure incorporated in it. In other
words, sampling cost should be directly proportional to length of
the sampled fragment and inversely proportional to the restraint strictness.
Sampling accuracy should be directly proportional to restraint radius.

To check conformity with these expected traits, we carried out backbone-only
sampling of filtered fragments under positional restraint of $1$,
$2$ and \Ang{3} on phosphorous atom. Note that fragments may be
the entire chains or at either terminus of the \rna chain or in between,
hence this also tests corresponding sampling styles. All-atom sampling
exercises were carried out under the same restraints on $P$ atom
and baseplane restraint of \Ang{5} on bases. $10$ modelling attempts
were made in each sampling exercise. A modelling attempt fails if
it cannot produce a model in $5$ trials. Each trial uses backtracking,
i.e. if sampling fails at a nucleotide, it is restarted from a position
$3$ nucleotides before it in the sampling order. In all-atom sampling,
glycosidic linkage ($\chi$ dihedral) is sampled uniformly over the
entire range at $10^{o}$ intervals. van der Waals radii of base and
sugar atoms are reduced by $50\%$. Sampling performance is quantified
by measuring the average \rmsd of models and average time taken to
produce a model as functions of fragment size, restraint radius and
whether bases are modelled.

\begin{figure}

\caption{Variation in sampling time with RNA fragment size. The following
scatter plots (above backbone-only, below all-atom) indicate a linear
correlation between average sampling times and fragment sizes. Regression
coefficients of lines of best fit suggest that sampling time nearly
doubles from \Ang{3} to \Ang{2} and almost quadruples from \Ang{2}
to \Ang{1}. Similarly, all-atom sampling is roughly a magnitude costlier
than backbone-only case. Note that 3 outliers have not been considered
for the \Ang{1} backbone-only plot and 4 fragments did not yield
any model during \Ang{1} all-atom sampling.}

\begin{center}\includegraphics[%
  width=125mm]{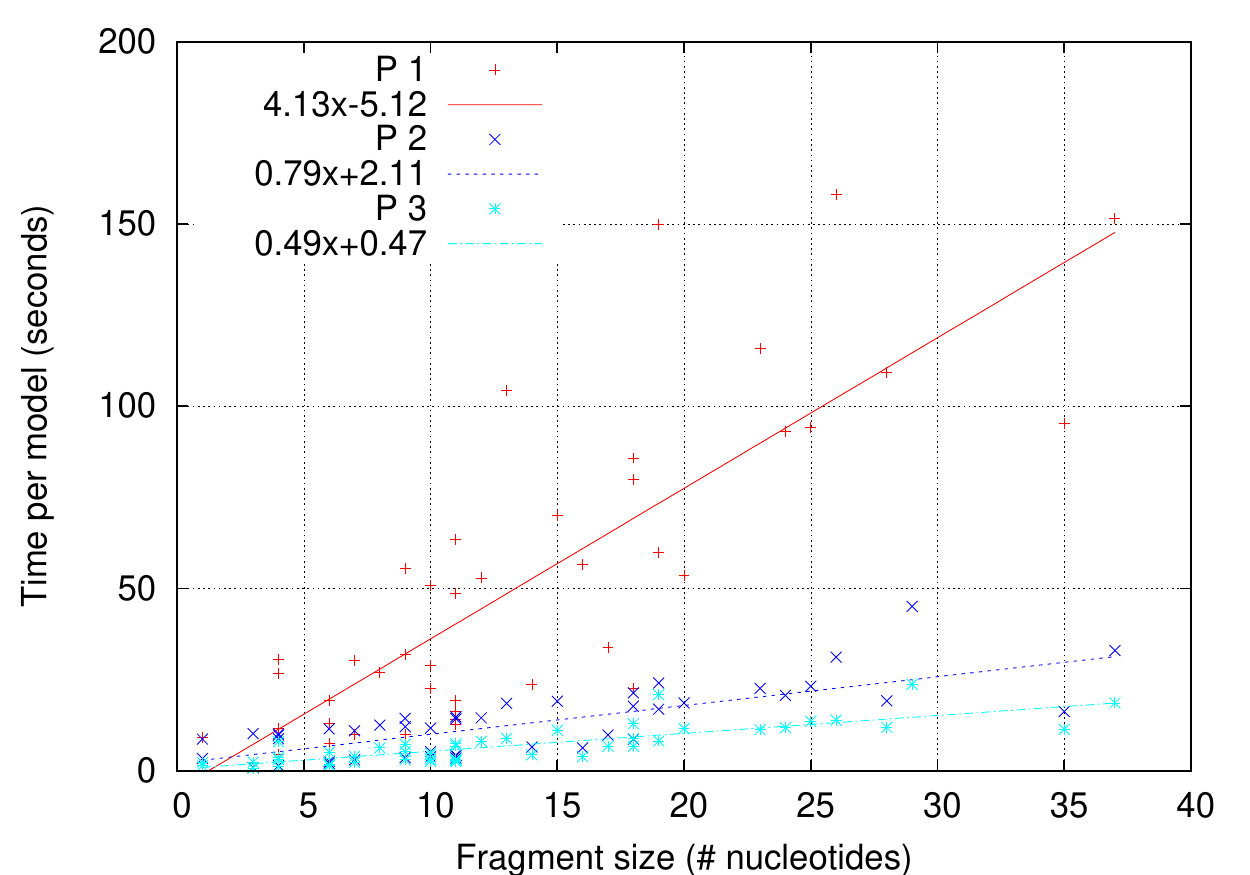}\end{center}

\begin{center}\includegraphics[%
  width=125mm]{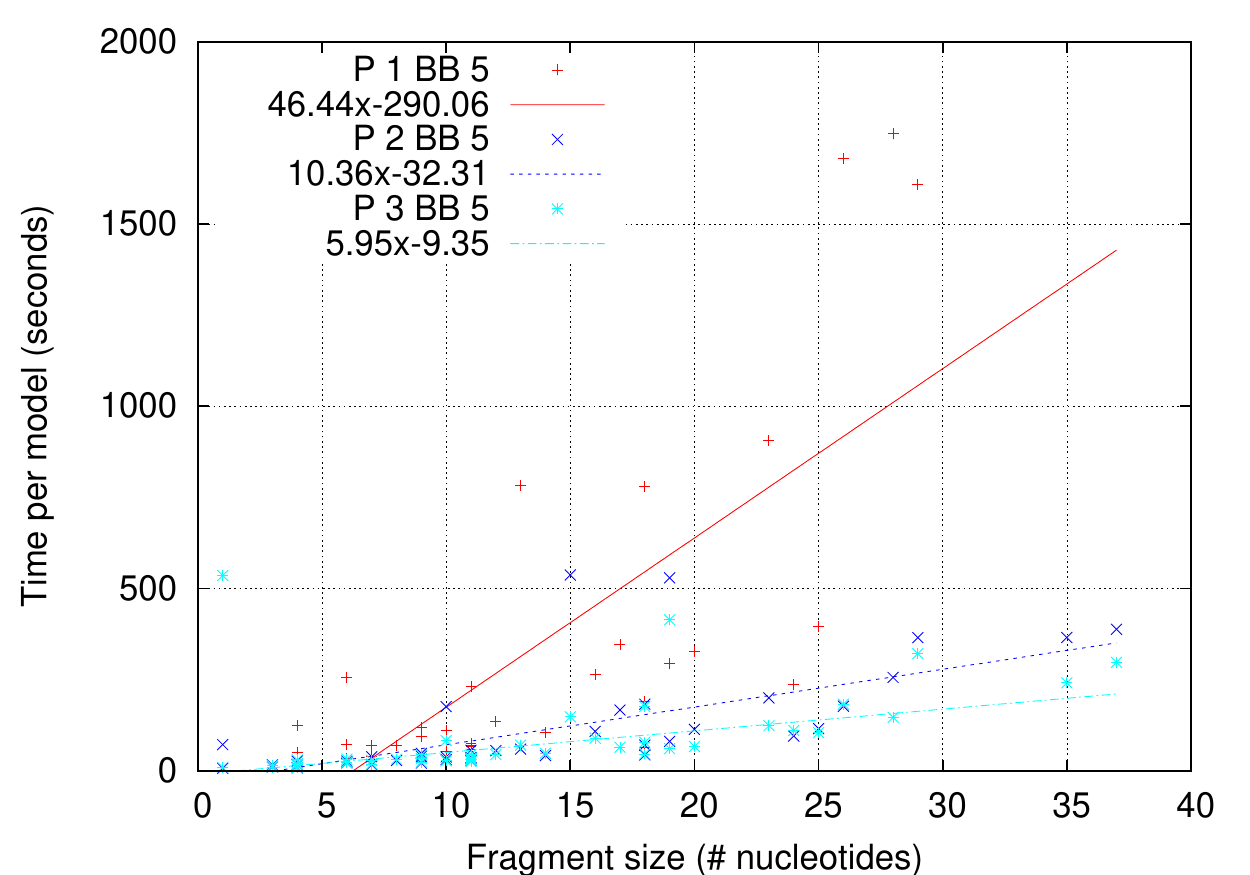}\end{center}

\label{samplingTime}
\end{figure}

The time plots (Fig.\ref{samplingTime}) suggest a linear correlation
between fragment size and sampling time for both backbone-only and
all-atom models, hence lines of best fit have also been plotted. Regression
coefficients of these lines are informative. In both cases, regression
coefficients suggest that the time needed for sampling with $P$ restraint
radius of \Ang{2} is twice as much as that with \Ang{3} and four
times as much with \Ang{1} as with \Ang{2}. Comparison of the regression
coefficients in backbone and all-atom cases suggest that latter is
nearly ten times costlier than the former.

\begin{figure}

\caption{Variation in sampling accuracy with RNA fragment size. The following
plots (top-\Ang{1}, middle-\Ang{2}, bottom-\Ang{3}) show relationship
between average \rmsd of models of fragments and fragment lengths.
There is a weak tendency to have lower \rmsd for lower lengths, hence
a log curve was fitted for each scatter. In general, at a given $P$
restraint radius, all-atom models have better backbone \rmsd than
backbone-only models. All-atom \rmsd is slightly greater than backbone
\rmsd in all-atom models. \rmsd increases as restraint radius increases.}

\begin{center}\includegraphics[%
  width=80mm]{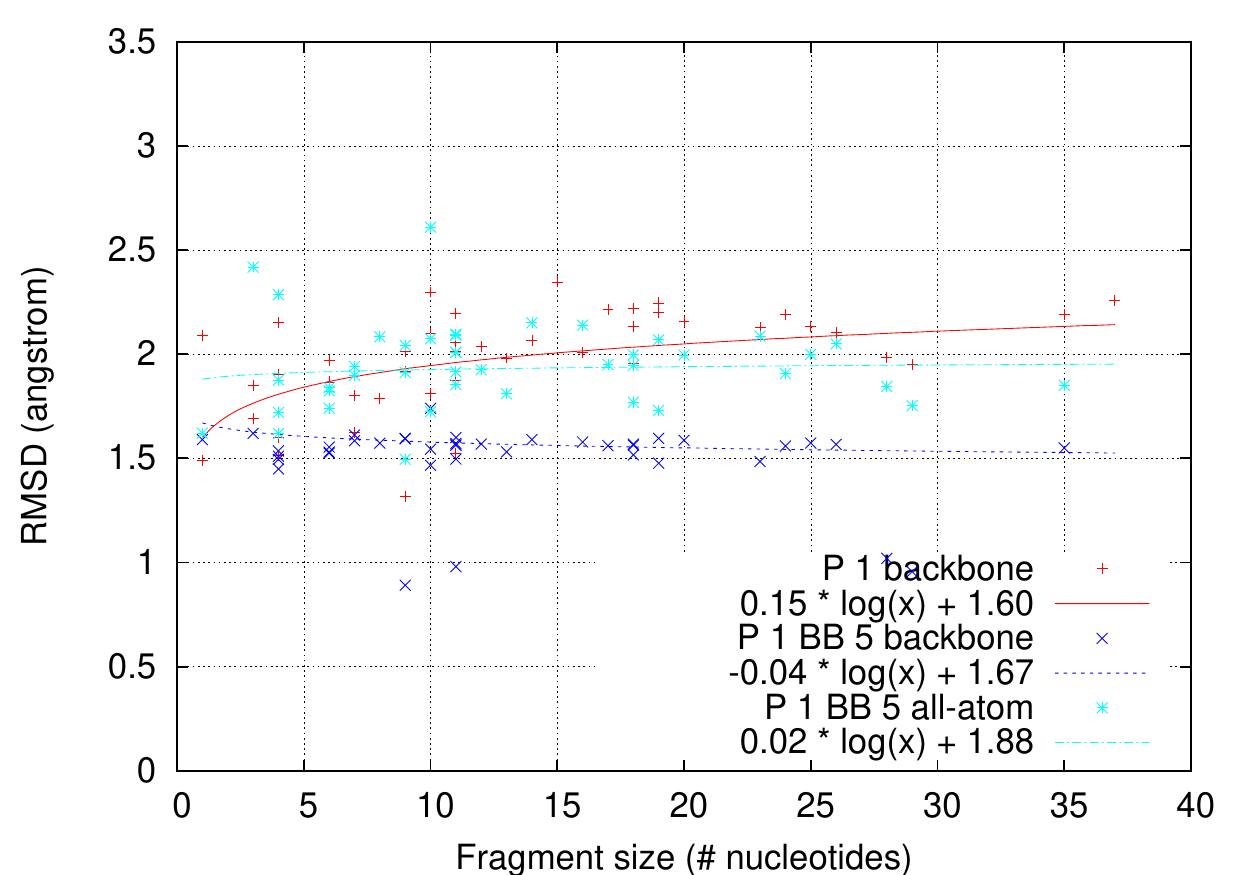}\end{center}

\begin{center}\includegraphics[%
  width=80mm]{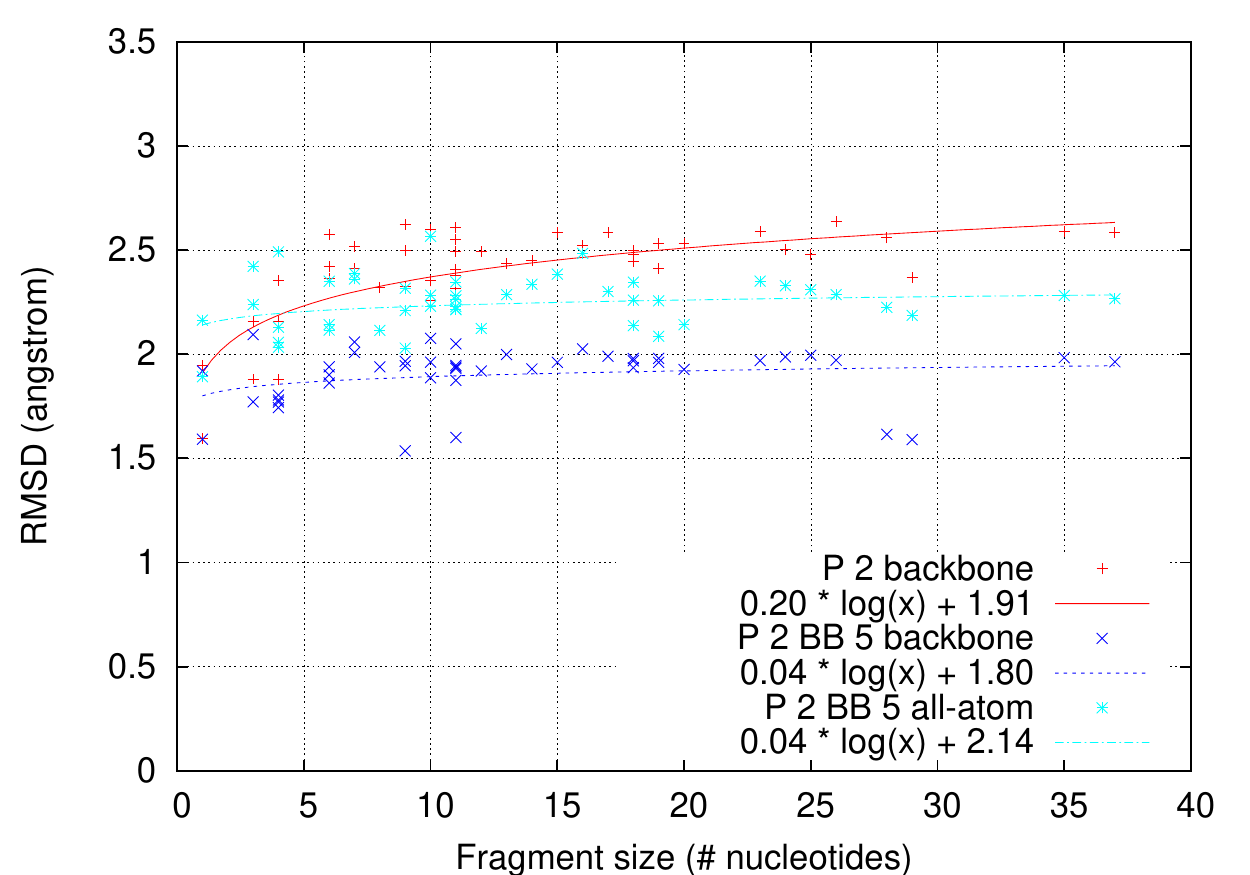}\end{center}

\begin{center}\includegraphics[%
  width=80mm]{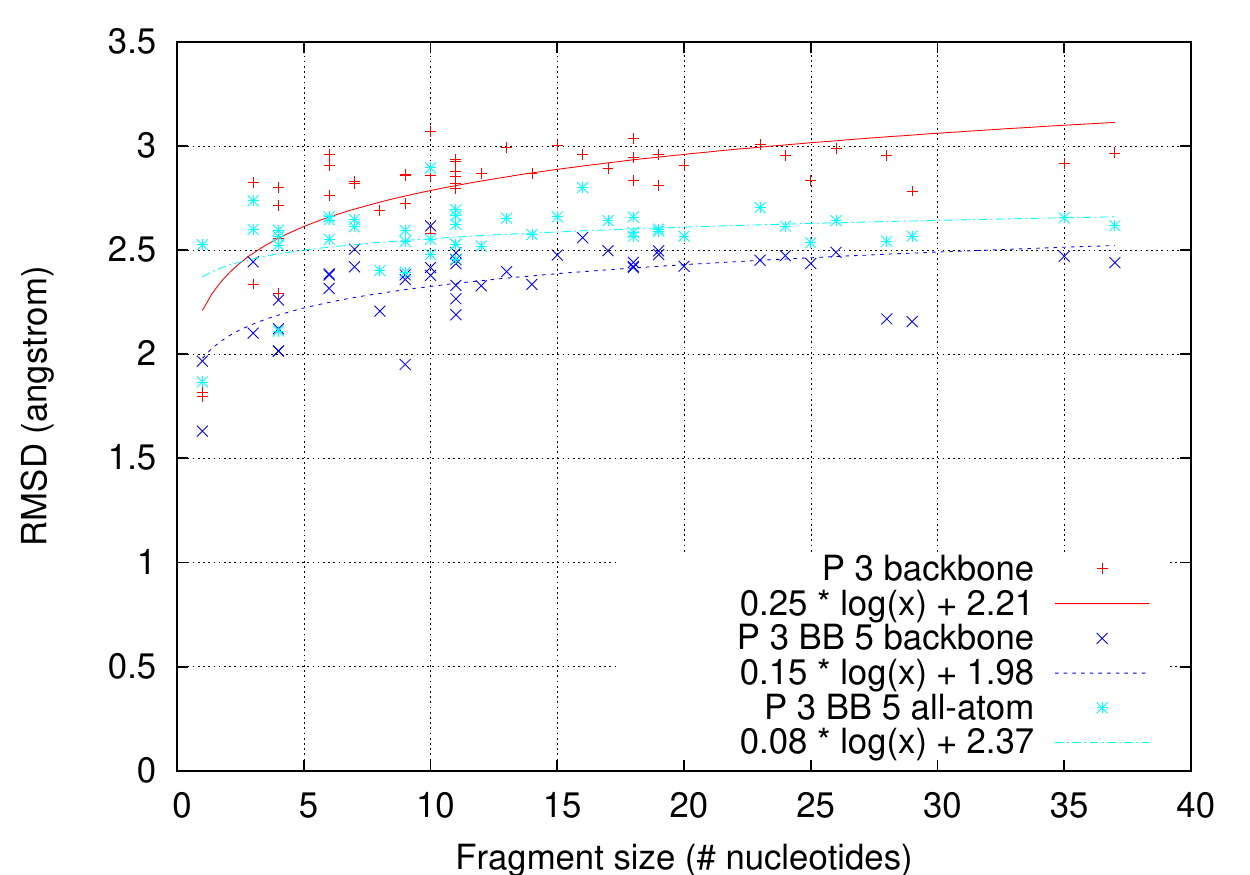}\end{center}

\label{samplingRMSD}
\end{figure}

The \rmsd plots (Fig.\ref{samplingRMSD}) suggest a weak correlation
between the \rmsd and fragment sizes, i.e. \rmsd is lower for smaller
fragments with the same restraint radius. This prompted the fitting
of a log curve. \rmsd falls with the size of $P$ restraint. For
each restraint size, all-atom \rmsd is more than backbone \rmsd.
Interestingly, the backbone \rmsd in all-atom case is better than
that in backbone-only case, indicating the influence of base restraint
in guiding the backbone.

Thus, the main sampling trends are: (a) smaller restraint radius leads
to greater sampling time, (b) all-atom sampling is costlier than backbone-only,
but leads to backbones with less \rmsd (c) sampling time is proportional
to fragment size and (d) \rmsd tends to be smaller for smaller fragments.
These trends are expected from previous experience with protein sampling
exercises. But there are significant differences too, due to differences
in restraint density. In protein $C_{\alpha}$ tracing, backbone sampling
models $3N$ atoms under $N$ positional restraints (ignoring carbonyl
oxygen), hence positional restraint density is $\frac{1}{3}$. In
the \rna backbone tracing exercise carried out here, this density
is $\frac{1}{6}$ (ignoring phosphate oxygens $O1P,O2P$). This is
reflected in the backbone \rmsd: in the case of proteins, backbone
\rmsd is generally lower than the $C_{\alpha}$ restraint radius
but it is generally higher for \rna than the $P$ restraint radius.
Another difference is rotamericity of protein sidechains and lack
of it in glycosidic linkages. This is indicated by lower all-atom
\rmsd for proteins than \rna chains under similar restraints. To
sum up, trends observed in \rna sampling are expected and satisfactory
enough to attempt application to the crystallographic scenario.

\section{Foray into crystallographic refinement}

\subsection{About t\rna structure}

\begin{figure}

\caption{tRNA structure: the schematic diagram shows the typical secondary
structure of tRNA. 3D representation below it shows all-atom and cartoon
representation of tRNA$^{Asp}$ as in PDB entry 2tra (\cite{2traPDB}).}

\begin{flushleft}\includegraphics[%
  width=80mm]{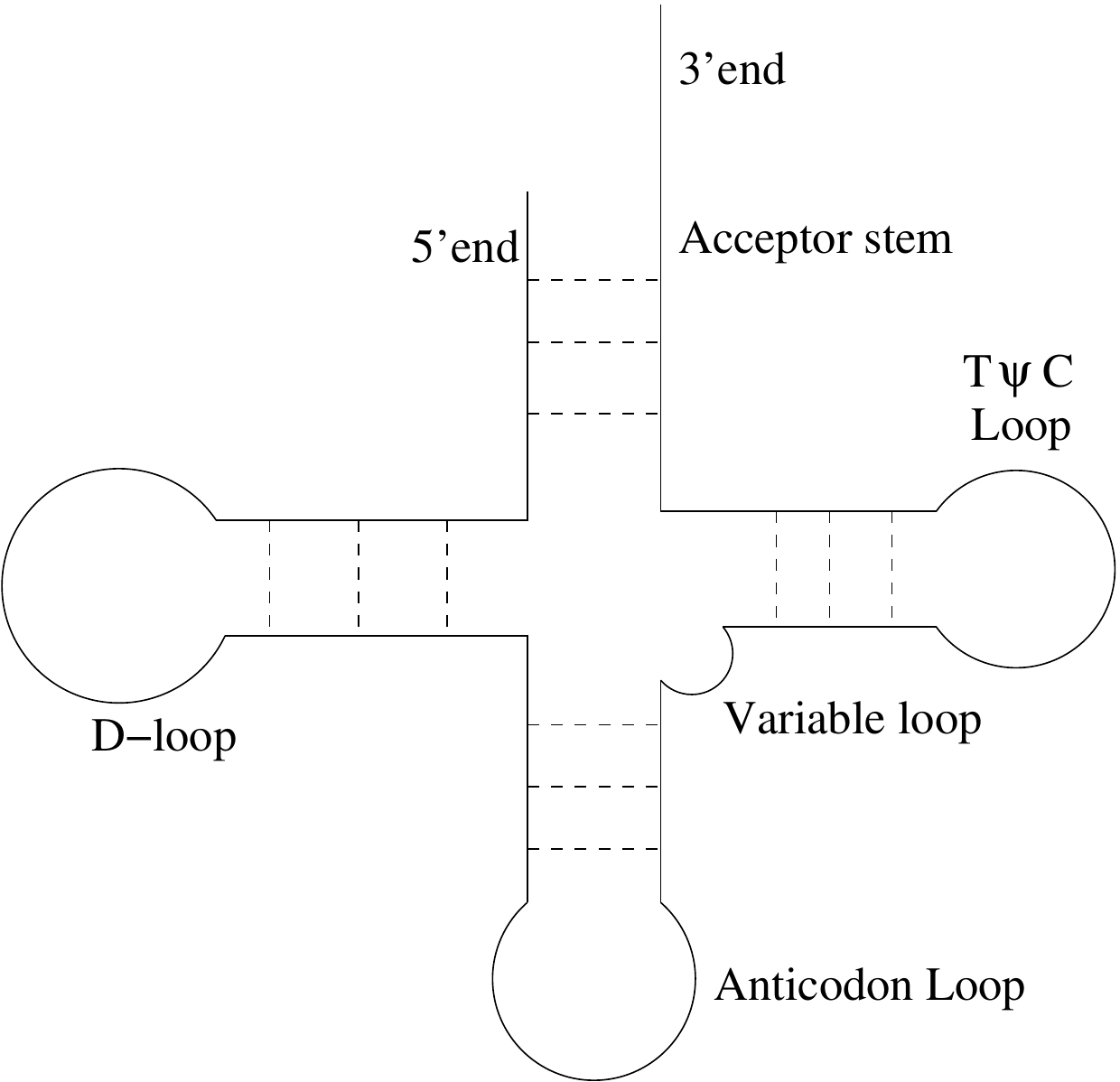}\end{flushleft}

\begin{flushright}\includegraphics[%
  width=80mm]{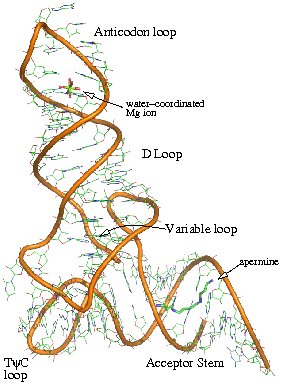}\end{flushright}

\label{tRNAstrs}
\end{figure}

Transfer \textsc{rna}s are classic structures from the 1970s. Till
mid-90s, structures of t\rna (\cite{4tnaPDB}, \cite{6tnaPDB}, \cite{2traPDB},
\cite{4traPDB}) were the only large \rna structures in \pdb (\cite{pheTRNArevisited}),
making them remarkable achievements of crystallography techniques
of that decade. t\rna is a cloverleaf-shaped molecule in its secondary
structure representation and has a L-shaped 3D form (Fig.\ref{tRNAstrs}).
t\rna is an essential cog in the translational machinery of the cell
which incrementally translates the transcripted m\rna into peptide
chain one residue at a time. Ribosome finds a t\rna with a 3-nucleotide
anticodon complementary to current m\rna codon. This t\rna has an
amino acid attached to its $5'$ end, which the ribosome then attaches
to the growing polypeptide.

t\rna structures are attractive for demonstrating crystallographic
utility of discrete restraint-based \rna sampling because they are
neither too small nor too large, are structurally well-studied and
have 3 loop regions (anticodon loop, $T\psi C$ loop, D loop) with
non-Watson-Crick base pairing. For this work, t\textsc{rna}$^{Asp}$
structure was used, solved at \Ang{3} by \cite{2traPDB}. This structure
(\pdb 2tra) refines to $R/R_{free}$ of $0.2552/0.3063$ with \cns
starting from deposited structure and data.

\subsection{Composite refinement protocol}

Similar to composite refinement protocols used earlier in the thesis,
this work also uses perturbed starting structures and rebuilds them
with the aim of improving $R_{free}$. In brief, \rappertk identifies
the ill-fit nucleotides by calculating the correlation coefficient
between $F_{c}$ map and $\sigma_{A}$-weighted \cns omit map for
regions around the backbone, sugar/base and entire nucleotide. Low
($<0.9$) correlation coefficient indicates nucleotide stretches to
rebuild, which are then built incrementally using \gabb algorithm.
Ten times more children are generated as the population size, and
top $10\%$ are retained based on their electron density occupation
score, leading to an enriched population. Resampled nucleotides get
a $B$-factor of $30$ assigned to all of their atoms. Non-\rna atoms
(ligands and waters) are not used during sampling. Best member of
population (according to density occupation) is written out as the
new model along with non-\rna atoms appended to it. The coordinates
and $B$-factors of non-\rna atoms are copied from the previous refinement
iteration. This model is refined with \cns (2 rounds of \mdsa starting
at $5000K$, intervened by a $200$-step minimization). This procedure
is repeated for $10$ iterations. It is expected that \rna models
generated with rotameric backbone states to obey given positional
restraints, positive electron density restraints and excluded volume
restraints would be within the convergence radius of \cns, i.e. such
models can be used to assist \cns in finding well-refined structures,
starting from ill-fitting ones.

\subsection{Refining a helical fragment}

\cns refinement was performed initially on the anticodon loop (nucleotides
$33-37$) and the $T\psi C$ loop (nucleotides $54-60$), starting
from models where the loops were perturbed by \rna tracing within
tight positional restraints ($P$, baseplane restraints of \Ang{2},
\Ang{5} respectively). In both cases we observed that \cns was able
to correct the errors introduced in the native structure. This was
in contrast to proteins where similar positional restraints on $C_{\alpha}$
and sidechains generally result in unsatisfactory \cns-only refinement.
But removal of baseplane restraints from the \rna trace deteriorated
the refinement quality. This suggested that \cns convergence radius
is larger for \rna structures than proteins, and \rappertk sampling
may be of value only in cases where spatial information about the
structure is highly uncertain.

\begin{figure}

\caption{CNS-only refinment of t\textsc{rna}$^{Asp}$ (PDB 2tra) can be unsatisfactory.
Starting models were generated by perturbing a 5 nucleotide fragment
(23-27) with \Ang{2} $P$ restraints and no base restraints. Top-left
panel shows the native structure of the fragment with its omit map
contoured at $2\sigma$. Other 3 panels show the best $R_{free}$
structures in 3 different CNS-only refinement trajectories, with respective
omit maps also contoured at $2\sigma$. This suggests that CNS can
get trapped in local minima in case of high initial structural uncertainty.
Note that the CNS refinement here is with minimal restraints, i.e.
hydrogen-bonding restraints between base-pairs were not provided to
CNS.}

\begin{center}\includegraphics[%
  width=60mm]{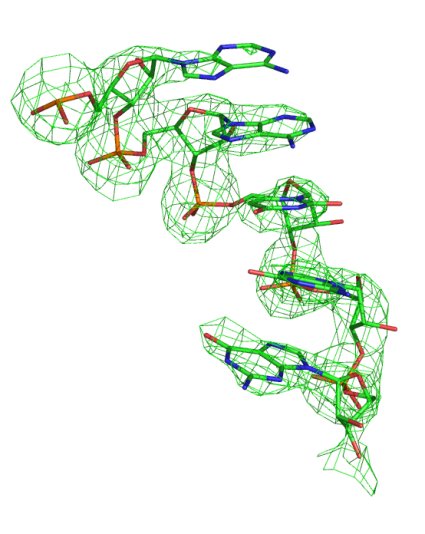}\includegraphics[%
  width=60mm]{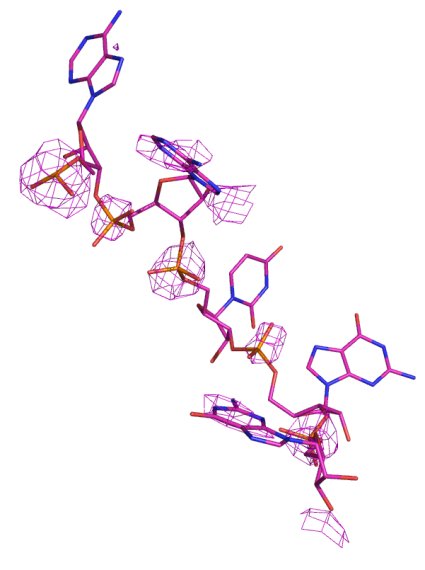}\end{center}

\begin{center}\includegraphics[%
  width=60mm]{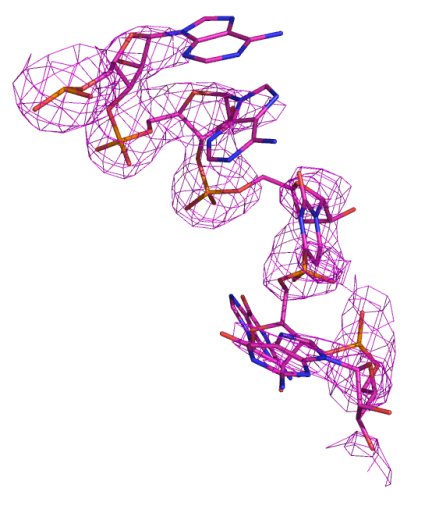}\includegraphics[%
  width=60mm]{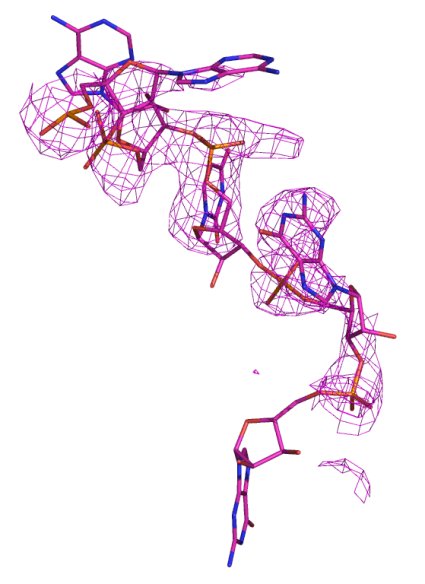}\end{center}

\label{RNAduplexCNStraps}
\end{figure}

\begin{figure}

\caption{Composite \cns/\rappertk refinement of a helical fragment from
t\textsc{rna}$^{Asp}$. Spherical positional restraints of radius
\Ang{2} were used around $P$ atoms of the 5-nucleotide (23-27) fragment
from PDB 2tra. No restraints were imposed on bases. The \cns/\rappertk
refinement resulted in satisfactory refinement in all $5$ attempts.
Best $R_{free}$ models in each trajectory are shown in magenta, with
the deposited structure in sticks representation.}

\begin{center}\includegraphics[%
  width=100mm]{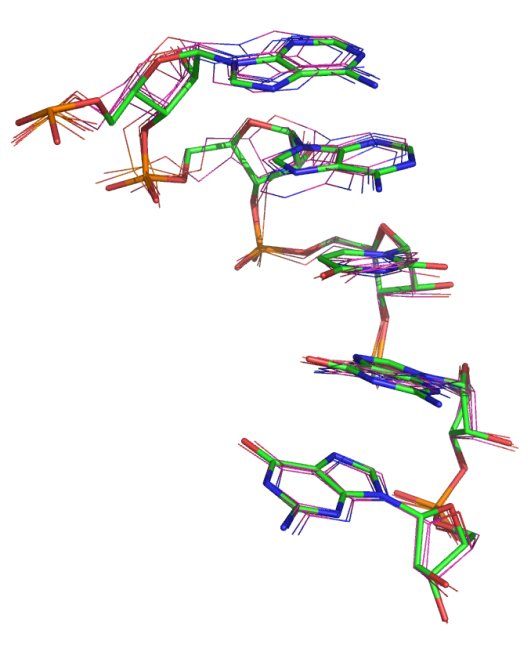}\end{center}

\label{RNAduplexRTK}
\end{figure}

In order to use a simple example to begin with, a fragment in \rna
duplex (nucleotides $23-27$) was chosen, with clear base densities.
Initial perturbation was carried out with \Ang{2} $P$ restraints
and no base restraints to generate $5$ models. The perturbed models
were subjected to \cns refinement only. In $4$ of $5$ cases, \cns
refinement was unsatisfactory. $3$ such cases are shown in Fig.\ref{RNAduplexCNStraps}.
When the composite refinement protocol was applied to the same region
with the same starting models, all trajectories resulted in well-refined
structures (Fig.\ref{RNAduplexRTK}). The mean of best $R_{free}$
values in \cns-only trajectories was $0.311$ as compared to $0.304$
for the composite protocol trajectories. It is interesting to note
that $R_{free}$ does not strongly reflect the salient differences
in the refinement trajectories indicated by Fig.\ref{RNAduplexCNStraps}
and Fig.\ref{RNAduplexRTK}.

\subsection{Refining the $T\psi C$ loop}

\begin{figure}

\caption{Composite \cns/\rappertk refinement of the 7-nucleotide $T\psi C$
loop (54-60) from t\textsc{rna}$^{Asp}$ with \Ang{2} $P$ restraints
and no base restraints. The top panel shows the native fragment with
its omit map density. Middle panel shows the best $R_{free}$ models
of \cns-only refinement. Bottom panel shows the same for the composite
refinement. Native structure (green) is shown for reference in middle
and bottom panels.}

\begin{center}\includegraphics[%
  width=75mm]{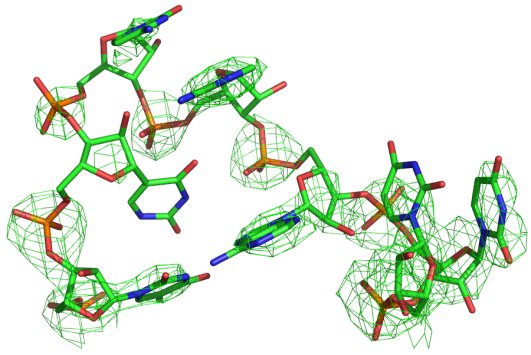}\end{center}

\begin{center}\includegraphics[%
  width=75mm]{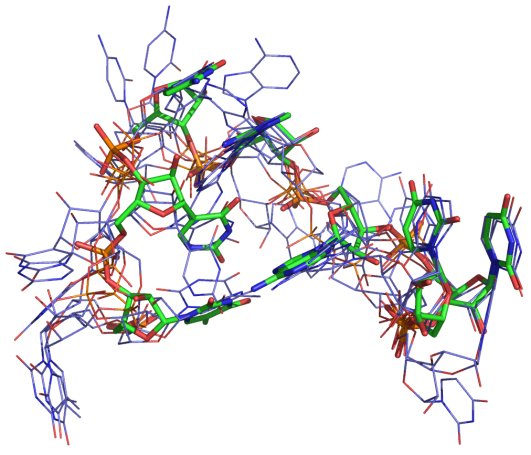}\end{center}

\begin{center}\includegraphics[%
  width=75mm]{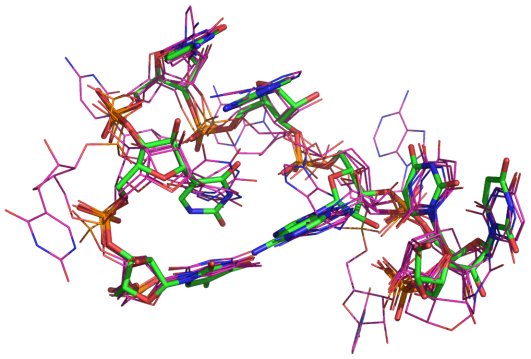}\end{center}

\label{TloopCNSRTK}
\end{figure}

The same exercise was repeated for nucleotides $54-60$, the $T\psi C$
loop. The native density for this loop is not as good as the helical
fragment (see Fig.\ref{TloopCNSRTK}). \cns-only refinement resulted
in mean best $R_{free}$ of $0.316$ over the $5$ refinement attempts,
whereas the same for composite refinement was $0.303$. Visual inspection
of these models shows the greater variability in the \cns models
and that each attempt was stuck in a local minimum. $3$ of $5$ composite
models refined to a structure very similar to native, but the rest
were trapped in a local minima. Close observation of these 2 cases
revealed that spurious density appearing elsewhere led \rappertk
sampling away from the native.

\subsection{Refining the anticodon loop}

\begin{figure}

\caption{Composite \cns/\rappertk refinement of the 5-nucleotide anticodon
loop (33-37) with \Ang{2} $P$ restraints and no base restraints.
The top panel shows the native fragment with its omit map density.
Middle panel shows the best $R_{free}$ models of \cns-only refinement.
Bottom panel shows the same for the composite refinement. Native structure
(green) is shown for reference in middle and bottom panels.}

\begin{center}\includegraphics[%
  width=100mm]{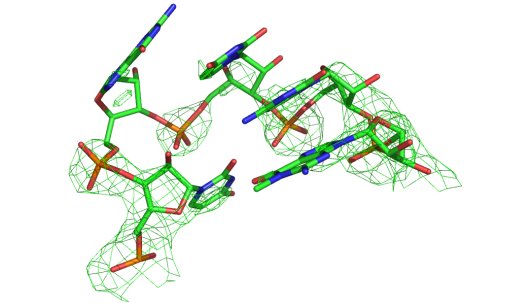}\end{center}

\begin{center}\includegraphics[%
  width=100mm]{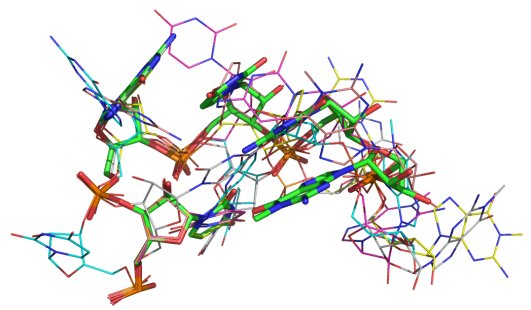}\end{center}

\begin{center}\includegraphics[%
  width=100mm]{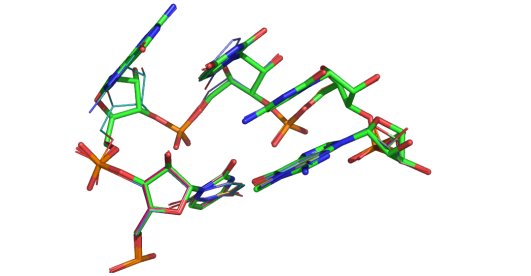}\end{center}

\label{AloopCNSRTK}
\end{figure}

Anticodon loop spans nucleotides $33-37$, of which $G-34$, $U-35$
and $C-36$ have two equally occupied states in the 2tra structure.
Initial attempts to repeat the previous exercise on this loop were
unsatisfactory because \rappertk tried to fit a single conformation
to these heterogenous nucleotides. Due to this, we created artificial
diffraction data at the same resolution by considering only the first
conformation of each nucleotide and assigning full occupancy to it.
This significantly changed the refinement trajectories and a similar
trend as previous two exercises could be observed. For five \cns-only
and composite refinements, mean best $R_{free}$ values were $0.254$
and $0.215$ respectively, indicating a much improved refinement with
the composite protocol. Fig.\ref{AloopCNSRTK} shows that the composite
protocol yields almost identical structures and \cns-only refinement
gets trapped in different local minima.

\subsection{Typical problems with refinement protocols}

\begin{figure}

\caption{CNS refinement of RNA can get trapped in local minima. An instance
of CNS-only refinement yields a structure (green) very different from
native (magenta). Corresponding $2F_{o}-F_{c}$ omit map is shown
contoured at $1\sigma$ around a three nucleotide stretch (green sticks,
nucleotides 54-56) only for clarity. Corresponding nucleotides in
the two structures are shown with arrows. CNS model has bases too
far away from native locations and also occupying the wrong density.
Also note the appearance of density around wrong bases.}

\begin{center}\includegraphics[%
  width=125mm]{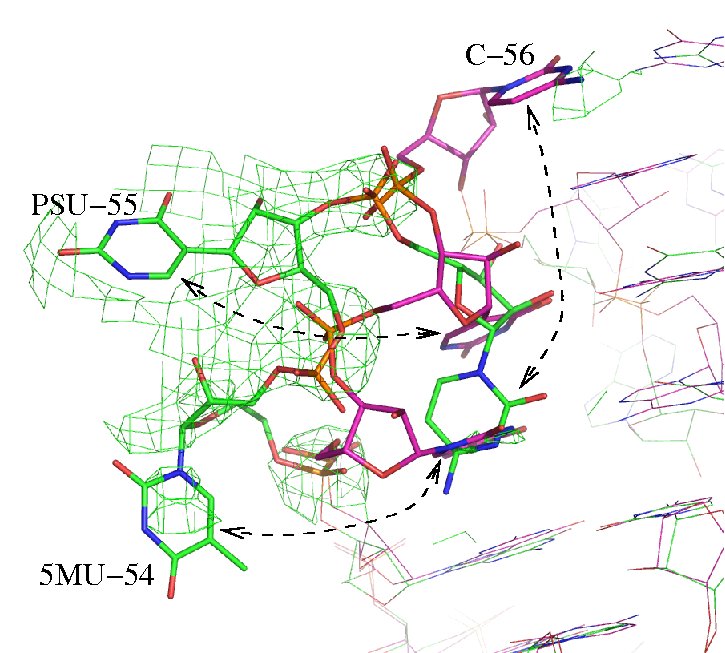}\end{center}

\label{CNSmistakes}
\end{figure}

There are two main reasons for suboptimal \cns-only refinement, identifiable
from successive structures in the refinement trajectories (Fig.\ref{CNSmistakes}).
Firstly, if a base is very far away from its native-like location,
\cns refinement does not restore it. Secondly, a base may get trapped
into densities of phosphate, sugar or another base, in which case
even if the base is not too far away, it is difficult to restore it.
This is reminiscent of bulky misplaced sidechains in protein crystallographic
refinement.

\begin{figure}

\caption{A typical corrective step carried out by \rappertk RNA sampling
with rotameric backbone and density enrichment. Blue is the native
structure, green is a perturbed and \cns-refined model. Magenta is
the \rappertk model found using the positional restraints and omit
map of the green model. Note that \rappertk model removes gross errors
yet small errors may remain in comparison to native.}

\begin{center}\includegraphics[%
  width=125mm]{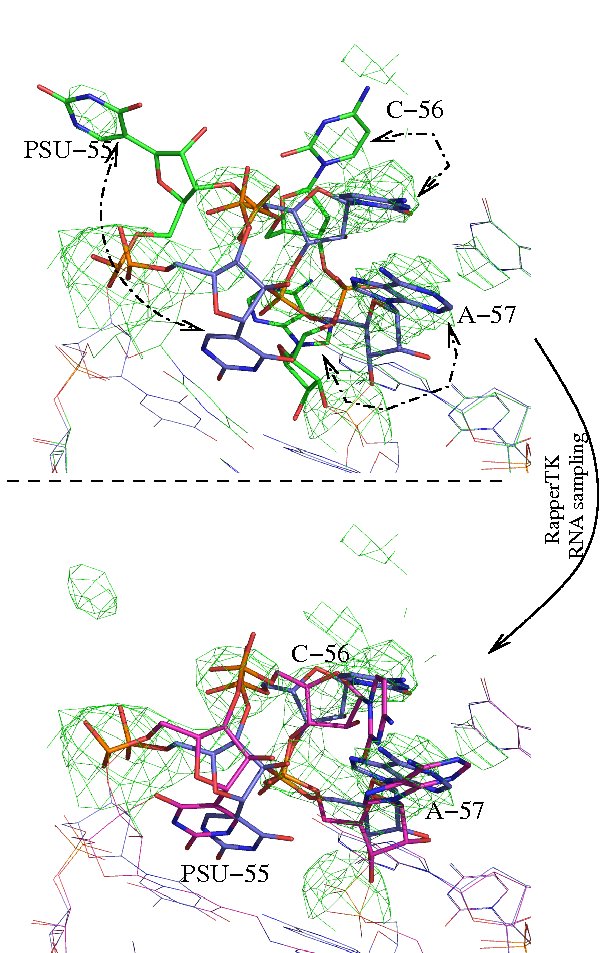}\end{center}

\label{beforeRTKafter}
\end{figure}

Structure trajectories suggest that improved refinement with \cns/\rappertk
protocol must be due to relocation of bases by \rna sampling, which
is brought about by the electron-density based enrichment of incremental
building using rotameric backbones. A typical corrective rebuilding
step is shown in Fig.\ref{beforeRTKafter}. The obvious mistakes in
base placement are corrected with \rappertk whereas \cns carries
out small corrections to take the conformations towards the optimal.

There are some imperfections in the \rappertk sampling scheme which
may sometimes lead to incorrect final structures: (a) \rna is very
flexible and population size of $300$ and enrichment factor of $10$
may not be sufficient (b) Lack of $\chi$ preferences means that selective
pressure due to bases is low - it is further weakened in case of weak
base density (c) Collateral damage may be caused by \cns refinement
of a defective loop, e.g. perturbations in nearby regions of structure
or symmetry-related copies do not get repaired during \rappertk rebuilding
step leading to higher $R_{free}$ (d) Scoring scheme based on maximizing
the electron density occupation may promote occupation of sharp peaks
like waters and phosphates although there are obvious dissimilarities
between such peaks and the shape of a base.

\section{Conclusion}

This work suggests that knowledge-based sampling can be applied efficiently
and productively to \rna structures. \gabb algorithm was extended
to sampling of \rna chains at $5'$-end, $3'$-end and intermediate
regions. Modified nucleotides were incorporated in addition to standard
ones for all-atom sampling. Using a 48-chains dataset, we showed that
sampling performance is along expected lines and suggests its suitability
for real-world applications like crystallography. Then we demonstrated
that a helical strand, the $T\psi C$ loop and the anticodon loop
in the t\textsc{rna}$^{Asp}$ structure can be automatically sampled
and iteratively refined using crystallographic data. It was found
that the composite \cns/\rappertk protocol yields structures better
refined than those by the \cns-only protocol. Shortcomings of both
protocols were discussed. This work shows that automated crystallographic
refinement of \rna chains is possible given the approximate trajectory
of phosphates. This is a promising result for reducing manual effort
and allowing exploration of multiple conformations.

Yet some concerns remain and must be addressed in future work. Sampling
preferences themselves are imperfect. A-form conformation is adopted
by more than $50\%$ \rna suites but population frequencies for rest
of the backbone rotamers are unclear. Hence we have used equal weights
for all suite rotamers. For similar reasons, we have not used the
weakly bimodal nature of the glycosidic linkage. A careful analysis
of available structural data will be required before incorporating
such preferences reliably, because sampling preferences are meant
to bias the conformational search and not restrict it. Another improvement
necessary for quicker sampling is the propagation of phosphate and
base restraint onto the backbone (e.g. on $C4'$) so that base restraint
satisfaction becomes more likely. At present this is a sampling bottleneck.

There are a few promising ways to extend this work. Firstly, whole-chain
crystallographic refinement of \rna structures can be performed for
low resolution structures to reduce the number of non-rotameric suites.
Secondly, \rna sampling can be used to generate 3D all-atom conformations
for secondary structures or motifs by expressing the base pairing/stacking
interactions as distance restraints. All conformations sampled to
satisfy these restraints will be useful in 3D structure prediction
which is, as noted before, a process of assembling 3D coordinates
of predicted secondary structures. Finally, protein and \rna sampling
can be combined together for automating the crystallography of protein-\rna
complexes, especially the very large ones like ribosomes so that human
attention will be required only in the early and late stages of refinement.

\bibliographystyle{marko}
\addcontentsline{toc}{section}{\refname}\bibliography{arxivRNA}

\end{document}